**Redefining Populations of Inference for Generalizations from Small Studies**

Wendy Chan, Jimin Oh, and Katherine J. Wilson

Graduate School of Education, University of Pennsylvania

March 07, 2022

**Abstract**

With the growth in experimental studies in education, policymakers and practitioners are interested in understanding not only what works, but for whom an intervention works. This interest in the generalizability of a study's findings has benefited from advances in statistical methods that aim to improve generalizations, particularly when the original study sample is not randomly selected. A challenge, however, is that generalizations are frequently based on small study samples. Limited data affects both the precision and bias of treatment impact estimates, calling into question the validity of generalizations. This study explores the extent to which redefining the inference population is a useful tool to improve generalizations from small studies. We discuss two main frameworks for redefining populations and apply the methods to an empirical example based on a completed cluster randomized trial in education. We discuss the implications of various methods to redefine the population and conclude with guidance and some recommendations for practitioners interested in using redefinition.

With the growth in randomized controlled trials in education, determining which interventions work and for whom has become a crucial question to policymakers and practitioners (Spybrook, 2014). Over the past decade, advances in statistical methods have contributed to the understanding of how the results from experimental studies may generalize, or apply, to target populations of inference, particularly when the study samples are not randomly selected (Stuart et al., 2011; Tipton & Olsen, 2018). These methods, primarily based on propensity scores, rely on the use of observable school characteristics to match and reweight study samples so that bias-reduced estimates of population average treatment effects can be derived (Stuart et al., 2011; Tipton, 2013a; O'Muircheartaigh & Hedges, 2014; Chan, 2017). Since their inception, propensity score-based methods have made important contributions in several areas of generalization research, including post-hoc analyses (Tipton, 2013a), study design (Tipton, 2013b), and longitudinal analyses (Chan et al., 2021).

While generalization research has advanced, an important challenge is that the sample sizes of many studies are small (Tipton et al., 2017). It is common for studies to include 30 – 70 schools, while generalizations are made to populations that are at least ten times larger. This is particularly the case when generalizations are not necessarily planned for in the study design so that a target population is not precisely defined a priori (Tipton et al., 2021). In this case, it is common to specify broad populations of inference (for example, all schools in the state) in post-hoc analyses. The challenge with small studies is that estimates of population average treatment effects can be imprecise (large standard errors) and biased. For the latter, bias arises when the schools in the (small) sample are systematically different from schools in the broad population on characteristics that potentially moderate treatment impacts.

In response to the challenges of generalizing from small studies, one suggestion that is made in practice is to redefine the inference population. Redefinition is the identification of a subset of the original population based on similarity to the study sample or based on a population of policy interest (such as all high poverty schools). The rationale for redefining the population is based on the idea that when inference is made on a subset of the original population, precision is expected to improve since the resulting subpopulation is smaller, and bias can potentially decrease if the subpopulation is more "like" the study sample. However, redefinition can be done in various ways and the choice of approach has implications on statistical inference for generalization.

This article seeks to highlight considerations for researchers, practitioners, and policymakers and offer guidance when redefinition is used to improve generalizations from small studies. We use "small" to refer to study samples that comprise less than 5% of the original inference population. We focus on the following research questions:

1. What are the implications on statistical inference when various approaches to redefining the population are used?

2. How should researchers and practitioners decide upon the appropriate redefinition method to use when generalizing from small studies?

We first give a brief review of propensity scores and introduce and describe two frameworks for redefining the population. We highlight their differences and discuss considerations in choosing one approach over the other. We then apply the approaches to an empirical example in education and discuss the implications for generalization research.

**Propensity Scores**

The strongest tool for generalization is probability or random sampling, but this type of sampling is rare in educational studies (Olsen et al., 2013). When the study sample is not randomly selected from an inference population, the concern is that systematic differences exist between schools that participate in a study and those that do not. If these differences moderate the schools' responses to an intervention, they can induce bias in estimates of the population average treatment effect (PATE; Imbens, 2004), making generalizations unwarranted. Statisticians developed propensity score methods to improve generalizations from experimental studies in the absence of random sampling. Propensity scores are the conditional probability of selecting into a study, modeled as a function of observable covariates or characteristics such as school size and urbanicity. The validity of propensity score methods for generalization depends on several assumptions. Among them is sampling ignorability, which states that the propensity score model must include all treatment effect moderators. Additionally, the distribution of covariates in the sample and population shares *common support* so that every school in the sample (population) has a comparison school in the population (sample) (Tipton, 2014). The notion of common support is crucial since generalizations to schools beyond the range of common support are generally based on extrapolation.

**Redefining Inference Populations**

When generalizations are based on small samples, the limited sample size poses two main challenges. First, small sample sizes are associated with larger standard errors, which affects the precision of PATE estimates. Second, because of the limited sample, the ratio of sample to population size is small, which affects the degree of common support (Chan et al., 2021). Because there are far fewer schools in the sample compared to the population, it is more likely that some population schools will be compositionally different (based on covariates) to sample schools.

Redefinition has been suggested as a potentially useful approach to improve precision and bias reduction. We argue that there are two main frameworks for redefining populations of inference and we refer to them as "quantitatively optimal" and "policy-relevant" approaches. In the following section, we describe each approach and discuss their advantages and limitations.

*Quantitatively Optimal Subpopulations*

Quantitatively optimal approaches are often based on propensity scores and they are optimal in the sense that the resulting subpopulations minimize the variance and bias of PATE estimates. In this study, we focus on three types of quantitatively optimal approaches. The first is based on a propensity score cutoff that was originally developed within the observational studies literature. The second is also based on propensity scores but uses a different set of cutoffs. The third and final approach uses the distributions of the covariates in place of propensity scores.

The first quantitatively optimal approach to redefine populations is based on the notion of "common overlap" in observational studies (Crump et al., 2009). Common overlap refers to compositional similarity in parts of the covariate distributions where the treatment and control group share similar values. When there is limited overlap between the two groups, conventional estimators of the average treatment effect can be significantly biased and imprecise (Dehejia & Wahba, 1999, 2002; Rosenbaum, 1989, 2001). Similar results have been found in generalization studies (Tipton, 2013a; Chan, 2021). Crump et al. (2009) proposed a method to redefine the original population using a cutoff value based on propensity scores. Using the optimal propensity score cutoff $\alpha$, observations that lay outside of the interval $[\alpha, 1 - \alpha]$ were discarded. The subpopulation in the range $[\alpha, 1 - \alpha]$ is considered "optimal" in terms of yielding an average treatment effect estimate with the smallest asymptotic variance. In practice, the range $[\alpha, 1 - \alpha]$ is well approximated by $[0.1, 0.9]$.

The second approach under the quantitatively optimal framework uses the range of the propensity scores to redefine the population (Potter, 1993; Scharfstein, Rotnitzky, & Robins, 1999; Lee, Lessler, & Stuart, 2011). Population schools whose propensity scores fall below the minimum or exceed the maximum propensity score in the sample are excluded. Variations of this approach based on the quantiles of the propensity score distribution have also been used. For example, Lee, Lessler, and Stuart (2011) used cut points ranging from the $50 – 99^{th}$ percentile of the propensity score range in the sample and excluded schools whose propensity scores lay above the cut points.

A third approach to redefining a population is to use the distributions of the covariates directly. In place of propensity score-based cutoffs, populations can be redefined using the ranges of the covariate distributions in the sample. Schools whose covariate values lie outside the range of the corresponding covariate distribution in the sample are excluded. For example, suppose school size was used as a covariate and schools in the sample had student populations that ranged from $500 – 1000$ students. Under the covariate-based approach to redefine a population, schools with student populations of less than $500$ or more than $1000$ would be excluded. This can be extended to multiple covariates in which population schools are excluded based on the collective distributions of the variables. An advantage of using covariates in place of propensity scores is that the redefinition does not depend on the correct specification of the propensity score model.

*Policy-Relevant Subpopulations*

Redefinitions can also be made informally. When the results of a study are meant to inform policy, another method to redefine populations of inference is to identify a subgroup of schools that may potentially be impacted by policies surrounding evaluation studies. These "policy-relevant" subpopulations can include schools that may be of concern to policymakers and practitioners (Stuart et al., 2016). Given the focus on equity in education, one policy-relevant subpopulation may be high poverty schools in which a significant proportion of students qualify for free- and reduced-priced lunch. Alternatively, another policy-relevant subpopulation may include low achieving schools whose average scores on an assessment lie in the bottom 25[th] percentile of the distribution. For studies that are funded by organizations such as the Institute of Education Sciences (IES), these policy-relevant subpopulations are often the focus of funded studies as policymakers are interested in interventions that best support disadvantaged students. An important feature of policy-relevant subpopulations is that they are often identified in an ad hoc manner; namely, identification is based on the general concerns and interests of various stakeholders and there is (typically) little consideration of formal statistical approaches to construct the subpopulation.

### *Considerations When Redefining Populations*

Because quantitatively optimal and policy-relevant approaches to redefining a population lead to various types of subpopulations, it is important to consider the implications of each method. The first consideration is the subpopulation size. If the original population was all public high schools in a state and the population was redefined to only include schools located in large metropolitan areas (what we refer to as a policy-relevant population), this redefinition may lead to a significant reduction in population size. While reducing the population size can improve the sample to population ratio and precision, prior studies have shown that some statistical estimators may still perform poorly, particularly if the sample contains fewer than 2% of the population (Schmid et al., 2020). Some quantitatively optimal approaches aim to preserve as much of the original population size as possible while identifying a subset that results in the smallest variance and bias of the average treatment effect. However, it is possible that a quantitatively optimal approach may include most of the original population, leading to a redefined population that is not much different from the original.

The second consideration is that a redefinition necessarily leads to changes in the causal estimate used for generalization. Because redefinition creates a new population that excludes schools from the original, estimates of the PATE may be different between the redefined and the original population, leading to different statistical inferences (Miratrix et al., 2021). For example, suppose a study was conducted that evaluated a reading curriculum designed to support literacy and reading comprehension among third grade students. Generalizing the effects of the reading curriculum to all schools in the state (a broad population) will likely lead to a different estimate compared to generalizing to all urban schools in the state (a subpopulation). These differences arise because of the potential heterogeneity in treatment response. Urban schools, for example, may include larger populations of English Language Learners (ELLs) and a reading curriculum that is effective in improving literacy may lead to larger effect sizes among urban schools compared to all schools in the state. In general, when researchers redefine the inference population, the causal estimate used in generalization studies will change when treatment effects are not constant and it is important to assess whether the new causal estimate aligns with the research questions of the study.

Finally, the third consideration is that redefining a population affects interpretability and optimality. Policy-relevant subpopulations may be easy to describe, but they may not be "quantitatively optimal" by statistical measures. That is, the schools included in policy-relevant subpopulations may not necessarily contribute to the minimization of the asymptotic variance of the average treatment effect. Similarly, while quantitatively optimal subpopulations may be optimal, they may be difficult to describe to a policy audience. This is also an important consideration for covariate-based approaches to redefine the population. While some studies have illustrated ways to improve the interpretability of optimal subpopulations, there are still tradeoffs between interpretability and optimality (Traskin & Small, 2011). In general, redefining an inference population for generalization can entail multiple changes to study and population parameters, and it is important to assess whether these changes align with the goals of the study.

**Motivating Example**

To contextualize the implications of redefinition, we introduce the motivating example, which is based on a completed cluster randomized trial (CRT) in education. We focus on this example as the original study size was small and the study was the focus of prior generalization research. From 2009 – 2010, the Indiana Department of Education and the Indiana State Board of Education implemented a new assessment system designed to measure annual student growth and to provide feedback to teachers (Konstantopoulos et al., 2013). In the CRT, 54 K-8 (elementary to middle) schools from Indiana volunteered to participate in the study to evaluate the effect of the assessment system on academic achievement. Of the sample of 54 schools, approximately half were randomized to implement the state's new assessment system (treatment) and the remaining schools were assigned to the business-as-usual condition. Among the treatment schools, students were given four diagnostic assessments that were aligned with the Indiana state test, and their teachers received online reports of their performance to dynamically guide their instruction in the period leading up to the state exam. The treatment impact was measured using the Indiana Statewide Testing for Educational Progress-Plus (ISTEP+) scores in English Language Arts (ELA) and mathematics.

Chan (2017) and Tipton et al. (2017) assessed the generalizability of the assessment system to a target population of 1,460 K-8 schools in Indiana during the 2009 – 2010 academic year. Both studies used propensity scores to match the sample of 54 schools to the target population on 14 covariates, which included continuous measures such as pretest scores, school size, attendance, as well as binary measures such as Title I status. Table 1 provides descriptive statistics of the covariates. The PATE was defined as the average difference in ISTEP+ scores in ELA and Math among schools that implemented the benchmark assessment system (treatment) and ones that did not (control). In both the original study and the generalization studies, the treatment impact of the benchmark assessment system was not statistically significant (Konstantopoulos et al., 2013; Chan, 2017; Tipton et al., 2017). Notably, the study sample comprised only 3% of the target population (54/1460) and this small sample to population size ratio was the focus of related generalization research on small samples (Tipton et al., 2017).

Table 1

In assessing the implications of the small sample size for generalization in the Indiana CRT, Tipton et al. (2017) noted that "sharp inferences" to large target populations were difficult and that reweighting approaches (like those based on propensity scores) were not always effective. Although redefinition was identified as a tool to improve inference in Tipton et al. (2017), it was not explored, leaving open the question of whether redefinition can be useful and what the implications are when different approaches are used.

**Redefinition and the Indiana CRT**

We examined eight redefinitions of the Indiana CRT population. Five were considered policy-relevant and they included the subpopulation of (i) urban schools, (ii) suburban schools, (iii) rural schools, (iv) high poverty schools in which over 75% of the student population qualify for free- and reduced-priced lunch and, (v) low achievement schools whose average ISTEP+ scores (Indiana's state assessment) for Math was in the lower 25[th] quantile of the distribution. These five examples of redefined populations are not exhaustive of all possible policy-relevant groups, but they represent the types of subpopulations that may potentially implement the intervention or be impacted by policies stemming from the results of the CRT. In addition, we also considered three quantitatively optimal subpopulations, which correspond to the approaches based on the propensity score cutoff in Crump et al. (2009), the cutoffs using the range of the sample propensity scores, and the covariate distributions. To be consistent with prior work (Chan, 2017; Tipton et al., 2017), we fit the same propensity score model using the covariates from Table 1 for the quantitatively optimal subpopulations.

**Estimators**

To estimate the PATE, we compared the performance of five estimators. In generalization studies, estimators of the PATE fall into three broad categories: (i) estimators that model sample selection (propensity scores), (ii) estimators that model the outcome directly and, (iii) estimators that model both sample selection and the outcome. We selected these five estimators for comparison because prior studies have shown that estimators such as Bayesian Additive Regression Trees (BART; Chipman, George, & McCulloch, 2007, 2010) and Targeted Maximum Likelihood Estimation (TMLE; van der Laan & Rubin, 2006) perform well in small samples (Schmid et al., 2020). Table 2 provides a brief description of each estimator. Because the choice of estimator is not the focus of this study, the descriptions are meant to provide a broad summary of each method. We refer the reader to the associated references for additional detail.

Table 2

**Effects of Redefinition**

In this section, we describe the effects of redefinition by focusing on three trends: (1) the effect on population size, (2) the effect on standard errors (precision) and (3) the effect on bias reduction. Throughout, we identify the subpopulations that were associated with the largest changes in the population parameters and highlight important differences between the quantitatively optimal and policy-relevant subpopulations. We first begin with Table 3, which provides the PATE estimates for the original population of $N = 1,460$ schools. Table 3 illustrates that there are notable differences among the estimates and the standard errors for the five estimators for both outcomes. In both ELA and Math, the IPW and EBLUP estimates and their standard errors were similar in magnitude, while the TMLE estimate had the largest standard error and was the only negative estimate among the methods. However, despite the differences in values, none of the estimates were statistically significant, which is consistent with prior studies (Chan, 2017; Tipton et al., 2017).

Table 4 provides the estimates for each of the redefined populations. Redefinition had a clear effect on population size ($N_0$). Most of the policy-relevant subpopulations, such as the subgroup of urban and suburban schools, had population sizes that were a fifth of the original population size $N$. The smallest subgroup was the suburban schools, which had a population size of 352 out of the original $N = 1,460$. In contrast, the PS Min/Max approach preserved most of the original population size, where redefinition only led to a 12% reduction. Interestingly, the Crump and covariates approach had nearly identical population size reductions and both subpopulations were comparable in size to the rural subpopulation.

Overall, redefinition improved (reduced) the standard errors in some cases, but the improvement was inconsistent across the estimators. The subpopulation in which precision improved for most of the estimators was the quantitatively optimal approach under Crump et al (2009). Under the Crump subpopulation, the standard errors of estimators like the IPW and outcome model were reduced by half compared to the original population and this was seen for both ELA and Math. However, the standard errors for BART experienced no change between the original and redefined population, which is potentially due to the extreme reductions in population size (Schmid et al., 2020). Among the remaining subpopulations, precision improved in the PS Min/Max and Covariates populations (both quantitatively optimal), and the Rural subgroup (a policy-relevant population), but the changes were inconsistent. The TMLE estimator, for example, had a larger standard error under the Rural subgroup for ELA while BART had the same standard error in both the Rural and original population. Notably, the Rural schools were the only policy-relevant subgroup in which precision improved for most of the estimators in both outcomes.

Table 3, 4

The redefinition analysis of the Indiana CRT suggests that precision improved in both quantitatively optimal and policy-relevant subpopulations, primarily in the Crump (quantitatively optimal) and Rural (policy-relevant) subpopulation. The final consideration is how redefinition potentially affected bias reduction. To assess the effects on bias, we measured the compositional similarity, based on observable covariates, between the study sample and schools in the redefined population using two statistical measures, the B-index (Tipton, 2014) and distributional overlap (Chan, 2021). The B-index, or generalizability index, is the geometric mean between the propensity score distributions of the sample and population. The B-index is bounded between 0 and 1, and values close to 1 indicate strong similarity or "high generalizability" between the sample and population while values below 0.50 indicate insufficient similarity (Tipton, 2014). Distributional overlap measures the proportion of population schools whose propensity scores fall in the range of the sample. Table 5 and Figure 1 provide the B-index and overlap for the original and the eight redefined populations.

<div align="center">Table 5, Figure 1</div>

Table 5 indicates that the compositional similarity between the Indiana CRT study and the original inference population ($N = 1,460$) is strong with a value of 0.90. Among the redefined populations, compositional similarity is lowest for the subgroups of high poverty and low achievement schools with B-index values of 0.39 and 0.30, which are associated with "low generalizability" (Tipton, 2014). As a result, schools in the CRT study sample are less "like" schools in the high poverty and low achievement populations in the state. The subpopulations based on the Crump, PS Min/Max and Covariates approaches (the quantitatively optimal approaches) to redefinition are considered even more similar compared to the original population. The similarity is strongest for the Crump subpopulation with a B-index of 0.98, the highest among the redefined populations. These trends are supported by Figure 1 where the high B-index values under the quantitatively optimal subgroups are consistent with the nearly complete overlap in propensity score distributions. Among the policy-relevant subgroups, compositional similarity is highest for the Rural and Suburban schools, but the similarity is not as strong compared to the quantitatively optimal subpopulations. Thus, if the covariates used in the propensity score model moderate the treatment impact of the Indiana assessment system, the bias of the PATE estimates is *smaller* in the Crump, PS Min/Max and Covariate subpopulations.

While changes to the PATE estimates were not a focus of the redefinition analysis, it should be noted that the estimates varied among the subpopulations. In some cases, the PATE estimates between the original population and subpopulation are notably different. The IPW estimate for ELA, for example, was 0.48 in the original population, but it is 1.19 in the subpopulation of high poverty schools (FRPL), which is indicative of the potential heterogeneity in treatment impacts across schools. These differences are also reflected in the differences in compositional similarity where the original study sample and the FRPL population appear to be significantly different (based on the covariates).

**Discussion of Implications**

The Indiana CRT analysis illustrates that redefinition can significantly affect precision, bias reduction, and population parameters such as population size and causal estimates. This study aimed to provide guidance for researchers and practitioners in deciding upon the appropriate redefinition approach to use. We believe that the choice of redefinition approach should be informed by the goal of the study. In the following section, we frame the discussion of implications around four main goals.

*Precision*

If the goal of redefinition is to improve precision alone, then both quantitatively optimal and policy-relevant approaches can be used. We see from the Indiana CRT study that standard errors for most estimators decreased in the Crump and Rural subpopulations. Note, however, that the extent of precision improvement may depend on the population size reduction since estimators like BART may not perform well with smaller population sizes. In this case, if the subpopulation is significantly smaller than the original population, this can offset any gains in precision from redefinition.

*Bias Reduction*

If the goal of redefinition is bias reduction alone, then the quantitatively optimal approaches are more appropriate. The B-index and overlap values increased under all quantitatively optimal subpopulations, which implied that the schools in these populations were more "like" the sample based on the observable covariates. Note, however, that there are two important considerations. The first is that compositional similarity is based on the observable covariates used in the study and the extent to which the given covariates are true moderators of treatment impacts. If the covariates are not relevant moderators, then it is uncertain how much bias reduction can be achieved with redefinition. The second is that while quantitatively optimal approaches are appropriate for bias reduction, they may not be easy to describe and may not include all schools of policy relevance. For example, the Crump subpopulation is based on propensity scores and since these are summary measures of multiple covariates, the resulting subpopulation features schools that were included based on a combination of covariate values (Rosenbaum & Rubin, 1983).

*Precision and Bias Reduction*

If the goal is to improve both precision and bias reduction, quantitatively optimal approaches are comparatively better. The approach under Crump led to consistent reductions in standard errors among all estimators and it was also the subpopulation that improved the most in compositional similarity (as assessed by the B-index and overlap). Because there are several quantitatively optimal approaches and likely others that are variations of the ones discussed, an important question is how to choose among them. One consideration is to examine the reduction in population size. Among the three approaches in this study, the PS Min/Max preserved most of the original population while improving precision and bias reduction overall. If the goal is to keep as much of the original population as possible, then the PS Min/Max approach may be the best compromise between quantitative optimality and population size. A second consideration is the interpretability of the resulting subpopulation. Both Crump and the PS Min/Max approach use propensity scores and the subpopulations will comprise a variety of schools. In this case, the Covariates approach may be better since the resulting subpopulation is defined by the covariates directly. However, it should be noted that interpretability may still be a challenge for the Covariates approach when the number of covariates used is large.

*Precision, Bias Reduction, and Policy Relevance*

Finally, if the goal of redefinition is to identify a subpopulation of policy relevance that can improve both bias reduction and precision, there may not be a single redefinition approach that is appropriate. While precision may improve in some policy-relevant subpopulations, bias reduction may not and the results from the B-index and overlap suggest that bias may actually *increase* if there is dissimilarity between the study sample and subpopulation. Alternatively, bias reduction and precision may improve under quantitatively optimal approaches like Crump et al. (2009), but the resulting subpopulation may be difficult to describe and may not necessarily include schools that are of policy relevance. In this case, it is important to consider the limitations of both the sample size and the sample itself. While small samples and the impacts on precision can be addressed using several redefinition approaches, bias reduction is dependent on the types of schools included in the sample. As a result, we recommend that researchers and practitioners make a choice about which factor(s) to prioritize when redefining a population since the limitations of sample size and sample composition will necessarily entail a compromise approach when it is difficult to achieve all goals.

**Conclusion**

When studies are based on small sample sizes, the validity of generalizations is often questioned. In this study, we considered the advantages and tradeoffs of redefinition as an approach to improve inference in small studies. We illustrated that redefinition takes many forms and the choice of method to use depends on the specific goals of the practitioners and policymakers. If, for example, the goal is to improve precision alone, both quantitatively optimal and policy relevant approaches can be used and the choice among these methods will have to be decided by the additional goals of the stakeholders.

While our recommendations are based on a single case study, we argue that the implications of redefinition are relevant to generalization studies that are based on small samples; namely, studies in which the ratio of sample to population size is small. Additionally, in many generalization studies in education, it is common to use the types of variables seen in the Common Core of Data or from state longitudinal data systems to construct the population data frames. As a result, the approaches to redefinition based on the covariates in the Indiana CRT are relevant to other studies that use similar data sources. Furthermore, our suggestions to practitioners and policymakers speak broadly to the considerations that should be made regarding bias reduction, precision, and policy relevance when redefinition is used and these considerations may be pertinent even when the study sample size is not small.

In sum, redefinition of the inference population offers possibilities in improving estimation, but it is not a method that will address all the limitations of small sample sizes. We conclude by noting that before redefinition is considered, it is important to return to one of the fundamental questions of generalization research, which is *for whom* are we interested in generalizing the results to? When variation in treatment impacts exist, addressing this question is crucial since a study's results will generalize differently to the original and to redefined populations. As policymakers and practitioners identify (clearly) the populations of inference that are of generalization interest, redefinition can be a useful tool in making precise and bias reduced inferences about the specific populations of concern.

## References

Austin, P. C., & Stuart, E. A. (2015). Moving towards best practice when using inverse probability of treatment weighting (IPTW) using the propensity score to estimate causal treatment effects in observational studies. *Statistics in Medicine*, *34*(28), 3661-3679.

Chan, W. (2017). Partially identified treatment effects for generalizability. *Journal of Research on Educational Effectiveness*, *10*(3), 646-669.

Chan, W. (2021). An evaluation of bounding approaches for generalization. *The Journal of Experimental Education*, *89*(4), 690-720.

Chan, W., Oh, J., & Luo, P. (2021). The Implications of Population Changes on Generalization and Study Design. *Journal of Research on Educational Effectiveness*, *14*(2), 471-500.

Chipman, H. A., George, E. I., & McCulloch, R. E. (2010). BART: Bayesian additive regression trees. *The Annals of Applied Statistics*, *4*(1), 266-298.

Chipman, H. A.; George, E. I.; and McCulloch, R. E. 2007. Bayesian ensemble learning. *Advances in Neural Information Processing Systems*, *19*, 265–272.

Cole, S. R., & Stuart, E. A. (2010). Generalizing evidence from randomized clinical trials to target populations: the ACTG 320 trial. *American journal of epidemiology*, *172*(1), 107-115.

Crump, R. K., Hotz, V. J., Imbens, G. W., & Mitnik, O. A. (2009). Dealing with limited overlap in estimation of average treatment effects. *Biometrika*, *96*(1), 187-199.

Dahabreh, I. J., Robertson, S. E., Steingrimsson, J. A., Stuart, E. A., & Hernan, M. A. (2020). Extending inferences from a randomized trial to a new target population. *Statistics in medicine*, *39*(14), 1999-2014.

Dehejia, R. H., & Wahba, S. (1999). Causal effects in nonexperimental studies: Reevaluating the evaluation of training programs. *Journal of the American Statistical Association*, *94*(448), 1053-1062.

Dehejia, R. H., & Wahba, S. (2002). Propensity score-matching methods for nonexperimental causal studies. *Review of Economics and statistics*, *84*(1), 151-161.

Hill, J. L. (2011). Bayesian nonparametric modeling for causal inference. *Journal of Computational and Graphical Statistics*, *20*(1), 217-240.

Imbens, G. W. (2004). Nonparametric estimation of average treatment effects under exogeneity: A review. *Review of Economics and Statistics*, *86*(1), 4-29.

Konstantopoulos, S., Miller, S. R., & van der Ploeg, A. (2013). The impact of Indiana's system of interim assessments on mathematics and reading achievement. *Educational Evaluation and Policy Analysis*, *35*, 481–499.

Lee, B. K., Lessler, J., & Stuart, E. A. (2011). Weight trimming and propensity score weighting. *PloS one*, *6*(3), e18174.

Miratrix, L. W., Weiss, M. J., & Henderson, B. (2021). An applied researcher's guide to estimating effects from multisite individually randomized trials: Estimands, estimators, and estimates. *Journal of Research on Educational Effectiveness*, *14*(1), 270-308.

Olsen, R. B., Orr, L. L., Bell, S. H., & Stuart, E. A. (2013). External validity in policy evaluations that choose sites purposively. *Journal of Policy Analysis and Management*, *32*, 107–121.

O'Muircheartaigh, C., & Hedges, L. V. (2014). Generalizing from unrepresentative experiments: a stratified propensity score approach. *Journal of the Royal Statistical Society: Series C (Applied Statistics)*, *63*(2), 195-210.

Potter, F. J. (1993, August). The effect of weight trimming on nonlinear survey estimates. In *Proceedings of the American Statistical Association, Section on Survey Research Methods* (Vol. 758763). Washington, DC: American Statistical Association.

Rao, J. N., & Molina, I. (2015). *Small area estimation*. New York, NY: John Wiley.

Rosenbaum, P. R. (1989). Optimal matching for observational studies. *Journal of the American Statistical Association*, *84*(408), 1024-1032.

Rosenbaum, P. R., & Rubin, D. B. (1983). The central role of the propensity score in observational studies for causal effects. *Biometrika*, *70*(1), 41-55.

Scharfstein, D. O., Rotnitzky, A., & Robins, J. M. (1999). Adjusting for nonignorable drop-out using semiparametric nonresponse models. *Journal of the American Statistical Association*, *94*(448), 1096-1120.

Schmid, I., Rudolph, K. E., Nguyen, T. Q., Hong, H., Seamans, M. J., Ackerman, B., & Stuart, E. A. (2020). Comparing the performance of statistical methods that generalize effect estimates from randomized controlled trials to much larger target populations. *Communications in Statistics-Simulation and Computation*, 1-23.

Spybrook, J. (2014). Detecting intervention effects across context: An examination of the precision of cluster randomized trials. *The Journal of Experimental Education*, *82*(3), 334-357.

Stuart, E. A., Cole, S. R., Bradshaw, C. P., & Leaf, P. J. (2011). The use of propensity scores to assess the generalizability of results from randomized trials. *Journal of the Royal Statistical Society: Series A (Statistics in Society)*, *174*(2), 369-386.

Stuart, E. A., Bell, S. H., Ebnesajjad, C., Olsen, R. B., & Orr, L. L. (2016). Characteristics of school districts that participate in rigorous national educational evaluations. *Journal of Research on Educational Effectiveness*, *10*, 168–206.

Tipton, E. (2013a). Improving generalizations from experiments using propensity score subclassification: Assumptions, properties, and contexts. *Journal of Educational and Behavioral Statistics*, *38*(3), 239-266.

Tipton, E. (2013b). Stratified sampling using cluster analysis: A sample selection strategy for improved generalizations from experiments. *Evaluation Review*, *37*(2), 109-139.

Tipton, E. (2014). How generalizable is your experiment? An index for comparing experimental samples and populations. *Journal of Educational and Behavioral Statistics*, *39*(6), 478-501.

Tipton, E., Hallberg, K., Hedges, L. V., & Chan, W. (2017). Implications of small samples for generalization: Adjustments and rules of thumb. *Evaluation Review*, *41*(5), 472-505.

Tipton, E., & Olsen, R. B. (2018). A review of statistical methods for generalizing from evaluations of educational interventions. *Educational Researcher*, *47*(8), 516-524.

Tipton, E., Spybrook, J., Fitzgerald, K. G., Wang, Q., & Davidson, C. (2021). Toward a System of

    Evidence for All: Current Practices and Future Opportunities in 37 Randomized Trials.

    *Educational Researcher*, *50*(3), 145-156.

Traskin, M., & Small, D. S. (2011). Defining the study population for an observational study to ensure

    sufficient overlap: a tree approach. *Statistics in Biosciences*, *3*(1), 94-118.

Van Der Laan, M. J., & Rubin, D. (2006). Targeted maximum likelihood learning. *The International

    Journal of Biostatistics*, *2*(1).

Table 1. Covariate Means for Indiana CRT and Indiana Population schools

| Covariate Descriptions | Indiana CRT $n = 54$ | Population $N = 1460$ |
| --- | --- | --- |
| 2008 - 2009 ELA Test Scores | 19.41 | 18.82 |
| 2008 - 2009 Math Test Scores | 16.48 | 16.46 |
| 2009 - 2009 Attendance | 96.44 | 96.38 |
| 2008 - 2009 Full Time Staff | 23.95 | 27.97 |
| 2008 - 2009 Population of Students | 423.07 | 480.98 |
| 2008 - 2009 Pupil-Teacher Ratio | 17.81 | 17.19 |
| County Population | 109878.63 | 230487.11 |
| 2008 - 2009 Title I Status | 0.83 | 0.77 |
| 2008 - 2009 Schoolwide Title I Status | 0.54 | 0.50 |
| Proportion of Male Students | 0.51 | 0.51 |
| Proportion of White Students | 0.84 | 0.77 |
| Proportion of Special Education Students | 0.17 | 0.16 |
| Proportion of Free/Reduced Price Students | 0.45 | 0.45 |
| Proportion of Limited English Proficiency Students | 0.02 | 0.04 |

Table 2. Estimators of the PATE

| Estimator | Type | Description |
|---|---|---|
| Odds Inverse Propensity Score Weighting (IPW) | Models sample selection | • IPW uses the inverse of the estimated propensity scores to create a weighted sample (Rosenbaum & Rubin, 1983; Austin & Stuart, 2015).<br>• PATE is estimated using a weighted difference in average outcomes or through weighted least squares regression (Cole & Stuart, 2010). |
| Bayesian Additive Regression Trees (BART) | Models outcome | • BART is a flexible modeling procedure that uses a sum of regression trees to model the outcome (Chipman, George, & McCulloch, 2010).<br>• PATE is estimated by taking the difference in outcomes between the treatment and control schools using repeated draws from a posterior distribution (Hill, 2011). |
| Outcome Modeling | Models outcome | • Uses a regression model based on the study sample to predict the treatment impact in the population<br>• PATE is estimated by averaging the predictions with respect to the covariate distributions in the population (Dahabreh et al., 2018). |
| Targeted Maximum Likelihood Estimation (TMLE) | Models both sample selection and outcome | • The TMLE uses a model to predict the outcomes under treatment and control for schools in the population and the predictions are adjusted using the propensity scores (van der Laan & Rubin, 2006). |
| Empirical Best Linear Unbiased Predictor (EBLUP) | Models outcome | • The EBLUP is a small area estimator (Rao & Molina, 2015) that uses administrative data to improve the precision of parameter estimates.<br>• PATE is estimated using a mixed effects linear model to predict the treatment impact (Chan, 2018). |

Table 3. PATE estimates for Indiana CRT (original population $P$)

| Estimator | ELA | Math |
|---|---|---|
| Odds IPW | 0.48 (0.34) | 0.72 (0.32) |
| BART | 0.17 (0.13) | 0.18 (0.15) |
| Outcome Model | 1.05 (0.37) | 1.64 (0.36) |
| TMLE | -0.01 (1.38) | 0.25 (1.89) |
| EBLUP | 0.35 (0.28) | 0.86 (0.32) |

Note: The original population comprised $N = 1{,}460$ schools. Standard errors are given in parentheses.

Table 4. PATE estimates for Redefined Populations in Indiana CRT

| Estimator | $P_0$: Rural ($N_0 = 856$) | | $P_0$: Suburban ($N_0 = 352$) | | $P_0$: Urban ($N_0 = 414$) | | $P_0$: FRPL ($N_0 = 421$) | |
|---|---|---|---|---|---|---|---|---|
| | ELA | Math | ELA | Math | ELA | Math | ELA | Math |
| Odds IPW | 0.17 (0.21) | 0.59 (0.15) | 0.39 (0.64) | 0.59 (0.71) | 0.46 (0.62) | 0.71 (0.71) | 1.19 (0.51) | 1.41 (0.20) |
| BART | 0.17 (0.13) | 0.17 (0.15) | 0.18 (0.13) | 0.18 (0.15) | 0.17 (0.14) | 0.20 (0.16) | 0.16 (0.15) | 0.20 (0.16) |
| Outcome Model | 0.86 (0.26) | 1.13 (0.26) | 1.45 (0.46) | 1.61 (0.42) | 1.17 (0.63) | 2.80 (0.60) | 1.17 (0.67) | 2.91 (0.60) |
| TMLE | 0.03 (2.08) | 0.20 (0.91) | 0.10 (0.40) | 0.24 (0.46) | 0.25 (0.19) | 0.97 (1.80) | 0.68 (3.07) | 0.76 (0.20) |
| EBLUP | 0.24 (0.20) | 0.59 (0.23) | 0.28 (0.30) | 0.75 (0.35) | 0.38 (0.33) | 0.90 (0.38) | 0.25 (0.15) | 0.26 (0.09) |
| | $P_0$: Math ($N_0 = 425$) | | $P_0$: Crump ($N_0 = 846$) | | $P_0$: PS Min/Max ($N_0 = 1335$) | | $P_0$: Covariates ($N_0 = 856$) | |
| Odds IPW | -0.84 (0.44) | 0.54 (0.61) | 0.31 (0.19) | 0.30 (0.22) | 0.42 (0.30) | 0.70 (0.28) | 0.11 (0.22) | 0.62 (0.19) |
| BART | 0.17 (0.14) | 0.19 (0.16) | 0.16 (0.13) | 0.14 (0.15) | 0.17 (0.13) | 0.18 (0.15) | 0.17 (0.13) | 0.17 (0.15) |
| Outcome Model | 1.12 (0.46) | 2.07 (0.48) | 0.59 (0.14) | 0.65 (0.14) | 0.90 (0.26) | 1.22 (0.25) | 0.73 (0.23) | 0.90 (0.20) |
| TMLE | -0.35 (0.41) | 0.30 (2.50) | 0.18 (0.56) | 0.09 (0.89) | -0.08 (3.99) | 0.24 (1.77) | -0.01 (1.38) | 0.25 (1.89) |
| EBLUP | -0.27 (0.43) | 0.55 (0.49) | 0.26 (0.16) | 0.19 (0.14) | 0.35 (0.28) | 0.86 (0.32) | 0.36 (0.17) | 0.32 (0.20) |

Note: The standard errors of each estimate are given in parentheses. The term $P_0$ refers to the redefined population and $N_0$ refers to the size of the redefined population. Rural refers to schools located in rural and town locales. Suburban and urban refer to schools in the respective locales. FRPL refers to high poverty schools in which over 75% of the student population qualify for free- and reduced-priced lunch. Math refers to low achievement schools whose average ISTEP+ scores for Math were in the lower $25^{th}$ quantile of the distribution. Crump refers to the subpopulation based on the propensity score cutoff in Crump et al. (2009). PS Min/Max refers to the subpopulation whose schools fall in the range of the propensity scores in the sample. Covariates refers to the subpopulation whose schools have covariate distributions that share the most overlap with those in the sample.

Table 5. B-index values for redefined populations in Indiana CRT

| Population | B-index |
|------------|---------|
| Original | 0.90 |
| Rural | 0.81 |
| Suburban | 0.80 |
| Urban | 0.58 |
| FRPL | 0.39 |
| Math | 0.30 |
| Crump | 0.98 |
| PS Min/Max | 0.92 |
| Covariates | 0.93 |

The B-index, or generalizability index, quantifies the compositional similarity, based on covariates, between the study sample and the given population. Values range from $0-1$, where larger values are associated with stronger similarity.

Figure 1. Distributional overlap in propensity scores for original and redefined populations in Indiana CRT

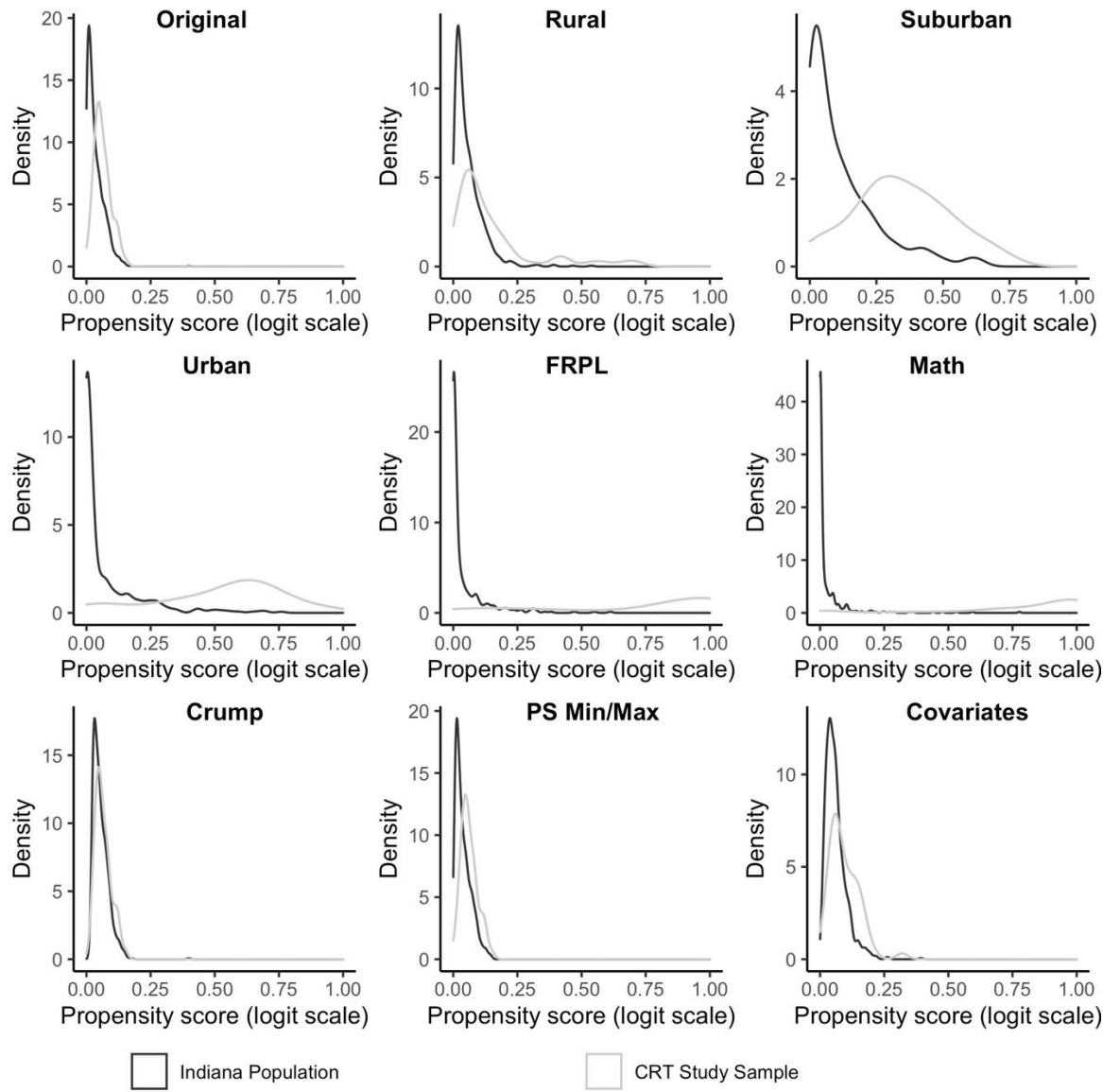

Indiana Population          CRT Study Sample